\documentclass{acm_proc_article-sp-tweeked}
\usepackage[pass,letterpaper]{geometry}
\usepackage{amsmath}
\usepackage{amsfonts}
\usepackage{txfonts}
\usepackage{mathrsfs}
\usepackage{listings}
\usepackage{tabularx}
\usepackage{graphicx}
\usepackage{subfigure}
\usepackage[usenames]{color}
\usepackage{ifpdf}
\usepackage{xspace}
\usepackage[draft]{changes}

\usepackage[pdfpagelabels=false]{hyperref}

\begin{document}

\markboth{S. Ainsworth & M. Nelson}{A Framework for Evaluation of Composite Memento Temporal Coherence}
\pagenumbering{arabic}


\title{A Framework for Evaluation of Composite Memento Temporal Coherence}

\author{
Scott G. Ainsworth\\
\small Old Dominion University\\
\small Norfolk, VA, USA\\
\small sainswor@cs.odu.edu
\and
Michael L. Nelson\\
\small Old Dominion University\\
\small Norfolk, VA, USA\\
\small mln@cs.odu.edu
\and
Herbert Van de Sompel\\
\small Los Alamos National Laboratory\\
\small Los Alamos, NM, USA\\
\small herbertv@lanl.gov
}
\maketitle


\begin{abstract}
Most archived HTML pages embed other web resources, such as images and
  stylesheets.
Playback of the archived web pages typically provides only the capture date
  (or \textrm{Memento-Datetime}) of the root resource and not the
  Memento-Datetime of the embedded resources.
In the course of our research, we have discovered that the Memento-Datetime of
  embedded resources can be up to several years in the future or past, relative
  to the Memento-Datetime of the embedding root resource.
We introduce a framework for assessing temporal coherence between a root
  resource and its embedded resource depending on Memento-Datetime,
  Last-Modified datetime, and entity body.
\end{abstract}

~

\section{Introduction}

Web archives, such as the Internet Archive \cite{carpenter-ndiipp10}, make
  best effort attempts to archive all or parts of the World Wide Web.
Although these archiving efforts are easily viewed as the virtual analog of
  a traditional library that collects books and periodicals, the differences
  between archiving the products of printing presses and archiving the pages
  of the Web are many.
This report focuses on the web archiving impact of a critical difference between
  composite physical and web products: when composition occurs.
When a book or magazine is produced, the text, images, and other resources
  are collected and composed into a product by the publisher.
Although future editions may use updated versions of the resources, a
  particular copy archived by a library is forever unchanging.
Future library patrons, when viewing the copy, need not consider if the copy
  currently in their hands changed after it was produced.
Thus, archiving physical media is deterministic from both the library's
  and patron's perspective.

In contrast to a book, web pages are composed much later in the production
  process: when it is viewed.
Returning to a library book, it is as if the book's photographs were
  not included until after the book is opened.
Web archives generally capture web resources using web crawlers such as
  Heritrix \cite{mohr-iwaw04}.
Due to resource constraints (e.g., network bandwidth) and ``politeness''
  considerations (not unduly burdening original resource servers), archive
  crawls seldom capture a web page and its component resources simultaneously.
Spaniol et al.\ \cite{spaniol-wicow09} note that crawls may span hours or days,
  increasing the risk of temporal incoherence because resources may change
  during the crawl.
Therefore, archiving composite web resources in a coherent state is
  probabilistic instead of deterministic.

\begin{figure}
  \centering
  \includegraphics[width=\columnwidth]{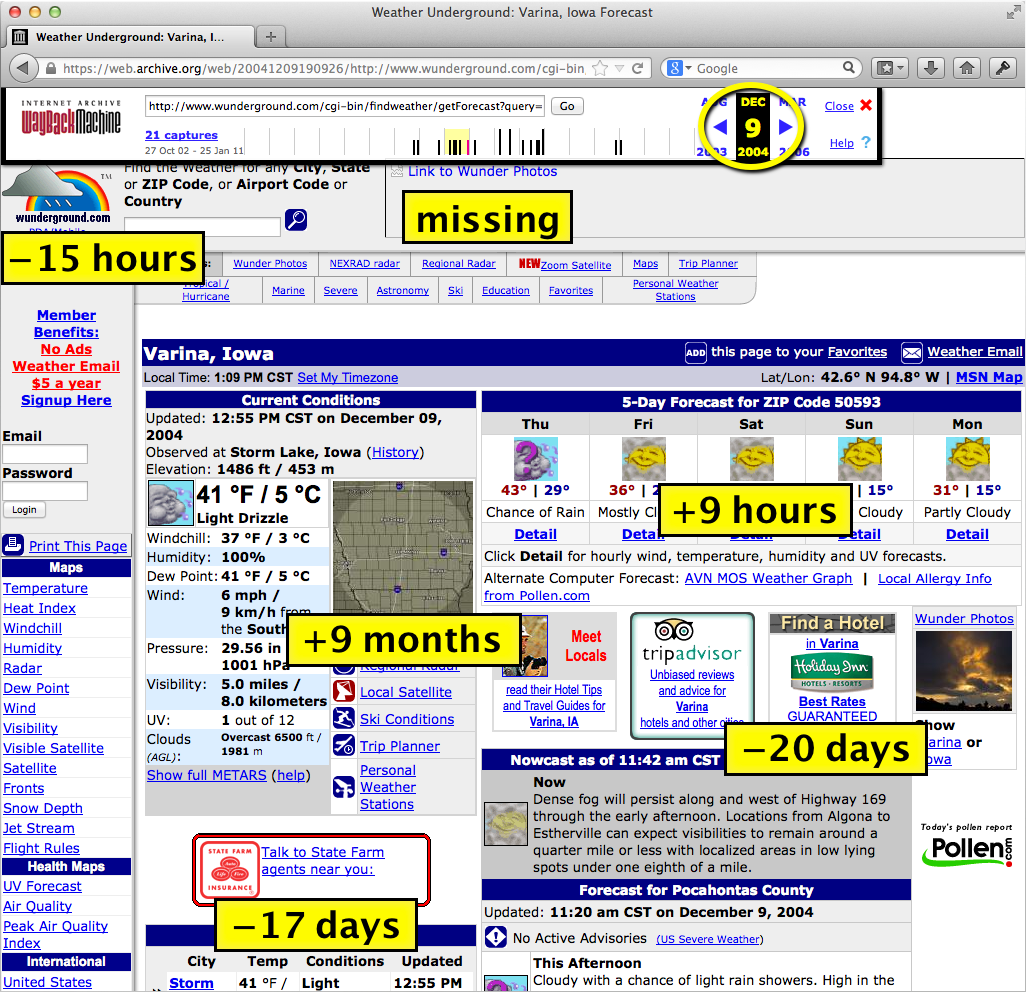}
  \caption{Embedded resource capture deltas, but entire page marked with a single datetime}
  \label{f:webpage:spread}
\end{figure}

Still, when an archived composite resource is presented in a web browser,
  it is labeled with the singular datetime of the root resource, as circled
  in figure \ref{f:webpage:spread}.
Most users would consider the presentation a coherent representation
  of the composite resource as it existed at that displayed datetime.
However, the capture datetime of the embedded resources can vary greatly.
Figure \ref{f:webpage:spread} also shows the capture deltas for five embedded
  resources for an archived wunderground.org forecast page that
  was captured by the Internet Archive on 2004-12-09 19:09:26Z.
The yellow rectangles indicate the capture deltas.
Negative means captured before the root; positive, after the root.
Note the \textit{clear weather} satellite image that was captured +9 months
  after the root.
Contrast it with the \textit{chance of rain} and \textit{mostly cloudy} images
  captured just 10 hours after the root.
This disagreement reveals a temporal incoherence---specifically, the
  satellite image is \textit{prima facie violative}.

In our research data, it is common for the temporal spread between the oldest
  and newest captures to be weeks or months.
Unexpected though was the discovery of many composite resources with spreads
  exceeding one year.
Even more unexpected were spreads exceeding five years; a few even exceed
  ten years.

Spaniol et al.\ \cite{spaniol-iwaw09}, Denev \cite{denev-vldb09},
  and Ben Saad et al.\ \cite{bensaad-tpdl11,bensaad-jcdl11} all introduce
  strategies to improve the quality of future web crawler-based captures.
If implemented, these strategies should reduce spread in new captures.
In contrast, our work focuses on the quality of existing
  holdings, without the benefit of improved crawl strategies.

Our investigation into the causes and consequences of temporal
  spread has revealed two important things about embedded resources:
\begin{itemize}
  \item Even if captured within seconds of the root resources, embedded
        resources are not always temporally coherent.
  \item Even if captured much later than the root resource, embedded
        resources are not necessarily incoherent.
\end{itemize}

Evaluation of temporal coherence depends on capture datetime,
  last modified datetime, and content differences.
Together, these attributes form patterns which can be considered prima facie
  coherent, prima facie violative, possibly coherent, and probably violative.
In order to facilitate reasoning about temporal coherence, we catalog embedded
  resource coherence patterns and the conditions that differentiate them.

The terminology used in this report comes primarily from the Memento framework
  \cite{vandesompel-arXiv:0911.1112,vandesompel-ldow10,rfc7089},
  which enables time-based HTTP access to archived resources.
The following definitions are from RFC 7089 \cite{rfc7089}:
\begin{description}
  \item[Original Resource]
    An Original Resource is a resource that exists or used to exist, and for
    which access to one of its prior states may be required.
  \item[Memento]
    A Memento for an Original Resource is a resource that encapsulates a prior
    state of the Original Resource.  A Memento for an Original Resource as it
    existed at time T is a resource that encapsulates the state the Original
    Resource had at time T.
  \item[Memento-Datetime] 
    The ``Memento-Datetime'' response header is used by a server to indicate
    that a response reflects a prior state of an Original Resource.  Its value
    expresses the datetime of that state.
  \item[TimeMap]
    A TimeMap for an Original Resource is a resource from which a list of URIs of
    Mementos of the Original Resource is available. 
  \item[URI-R]
    The URI of an Original Resource.
  \item[URI-M]
    The URI of a Memento.
  \item[URI-T]
    The URI of a TimeMap.
\end{description}
At first glance, the Memento-Datetime header appears to duplicate the
  Last-Modified header; this is not the case.
Last-Modified is set by an original resource's server and indicates the last
  datetime the resource changed.
Memento-Datetime is the datetime the original resource was captured and
  archived.
The existence of the \textrm{Memento-Datetime} header
  also entails a promise that the original resource's state
  is archived, frozen in time \cite{rfc7089}.
See the appendix for additional details and example headers.

\section{Composite Mementos}
\label{s:CompositeMementos}

\subsection{Definition}
\label{ss:Definition}

\begin{figure}
  \centering
  \includegraphics[width=\columnwidth]{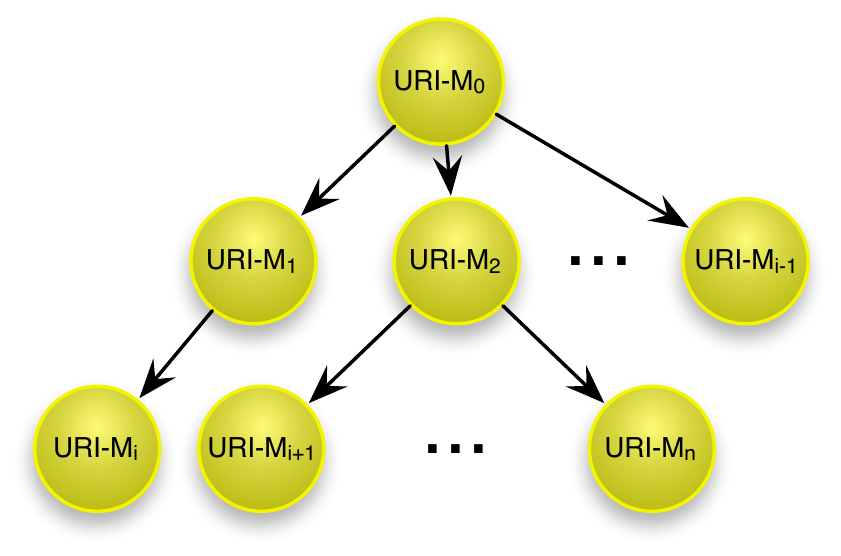}
  \caption{Composite Memento}
  \label{f:CompositeMementoTree}
\end{figure}

A \textit{composite memento} is a root URI-M and
  all embedded URI-Ms required to recompose the presentation
  at the client.
A composite memento generally comprises a root HTML resource and embedded
  resources such as images and stylesheets.
Some embedded resources (e.g., HTML frames, style sheets) can also have embedded
  resources, which are also part of the composite memento.
Figure \ref{f:CompositeMementoTree} is a tree representation of a composite
  memento.
(Technically, a composite memento is a graph, but in this context can treated as a
  tree without loss of generality.)

\subsection{Recomposition}
\label{ss:Recomosition}

\newcommand{\rurir}{\ensuremath{R_0}\xspace}
\newcommand{\eurir}[1]{\ensuremath{R_{#1}}\xspace}
\newcommand{\rurim}{\ensuremath{M_0}\xspace}
\newcommand{\eurim}[1]{\ensuremath{M_{i,#1}}\xspace}
\newcommand{\rmem}{\ensuremath{m_0}\xspace}
\newcommand{\emem}[1]{\ensuremath{m_{i,#1}}\xspace}

Recomposing a composite memento is the recursive process of selecting
  URI-Ms for URI-Rs, retrieving the representations for the
  URI-Ms, and extracting URI-Rs embedded in those URI-M representations.
The variables in table \ref{t:definitions} and functions
  in table \ref{t:functions} are used in the description that follows.
$R$ is simply a short form for URI-R, with a root URI-R designated \rurir
  and embedded URI-Rs designated \eurir{i}, $i > 0$.
Likewise, $M$ is short for URI-M.
The representation obtained by dereferencing $M$ is $m$.
A timemap, $\mathbb{T}$, for an $R$ is a set of $M$, which can be empty.
A composite memento, $\mathbb{C}$, is a root URI-M, \rurim, and zero
  or more embedded URI-Ms, \eurim{j}, one for each \eurir{i} (images,
  css, etc.).

\begin{table}[t]
\centering
\caption{Definitions}
\label{t:definitions}
\renewcommand{\tabcolsep}{1mm}
\begin{tabularx}{\linewidth}{cX}
  Term  & Definition \\
  \hline
  $R$
      & An original resource URI.  A URI-R. 
        \rurir is the root URI-R.
        $R_i$, $i\ge1$, is an embedded URI-R. \\
  $M$
      & A URI for an archived copy of a URI-R. A memento URI. A URI-M.
        \rurim is the root URI-M.
        \eurim{j}, $1 \le j \le n$, is a URI-M for \eurir{i}. \\
  $m$
      & The representation of $M$.
        \rmem is the representation of \rurim.
        \emem{j} is the representation of \eurim{j}. \\
  $\mathbb{T}$
      & A timemap for $R$. A list of mementos for $R$.
        $\mathbb{T}_i = \{ \eurim{j} | 1 \le j \le n \}$; $n$ is the number of $M$s
        archived for \eurir{i}. \\
  $\mathbb{C}$
      & A composite memento. A set of $M$ comprising the root URI-M, \rurim,
        and zero or more embedded URI-Ms, \eurim{j}. \\
  $t$
      & A target datetime (\textit{Accept-Datetime} in RFC 7089 \cite{rfc7089}). \\
  \hline
\end{tabularx}
\end{table}

\begin{table}[t]
\centering
\caption{Functions}
\label{t:functions}
\renewcommand{\tabcolsep}{1mm}
\begin{tabularx}{\linewidth}{lX}
  Function & Definition \\
  \hline
  $\mathcal{E}(m)$
      & Set of $R$s embedded in $m$. \\
  $\mathcal{H}(R,t)$
      & Heuristic providing the best $M$ for $R$ at $t$. \\
  $\mathcal{G}(M)$
      & Representation of $M$; generally retrieved via HTTP GET. \\
  $\mathcal{C}(R,t,\mathcal{H})$
      & Composite memento for $R$. \\
  \hline
  \multicolumn{2}{l}{\footnotesize \textsuperscript{\textdagger}See
                     RFC 2616 \cite{rfc2616}.}
\end{tabularx}
\end{table}

The recomposition process is represented by algorithm \eqref{e:recompose}.
Three parameters are required: $R$ is a URI-R, $t$ is a target datetime,
  and $\mathcal{H}$ is a memento selection heuristic.
The algorithm states that a composite memento is the set of recursively
  selected URI-Ms, one per URI-R, starting with URI-R, $R$, for
  target-datetime $t$, using heuristic $\mathcal{H}$.
Heuristic $\mathcal{H}$ is a function that returns the best\footnote{The
  definition of \textit{best} depends on the heuristic. A common definition is
  least difference between capture datetime and target datetime.
  RFC 7089 \cite{rfc7089} does not specify a memento heuristic.}
  $M$ for $R$ at target datetime $t$.
Thus, $\mathcal{G}(\mathcal{H}(R,t))$ is the best memento for $R$ at
  $t$ under $\mathcal{H}$.
The rest of the algorithm, $\{ \mathcal{C}(R_i,t,\mathcal{H}) \ |\ 
  R_i \in \mathcal{E}(\mathcal{G}(\mathcal{H}(R,t)))\}$,
  is the set of mementos for the URI-Rs embedded in $R$, applied recursively.
\begin{equation}
  \mathcal{C}(R,t,\mathcal{H}) = 
    \mathcal{G}(\mathcal{H}(R,t)) \cup
    \{ \mathcal{C}(R_i,t,\mathcal{H}) \ |\ R_i \in \mathcal{E}(\mathcal{G}(\mathcal{H}(R,t))) \}
  \label{e:recompose}
\end{equation}
Given algorithm \eqref{e:recompose}, the composite memento, $\mathbb{C}$, for root
  URI-R, \rurir, for target datetime $t$, under heuristic $\mathcal{H}$
  is shown in \eqref{e:composite:memento}.
\begin{equation}
  \mathbb{C} = \mathcal{C}(\rurir,t,\mathcal{H})
  \label{e:composite:memento}
\end{equation}

\subsection{Temporal Spread}
\label{ss:TemporalSpread}

\begin{figure}[h!]
  \centering
  \includegraphics[width=\columnwidth]{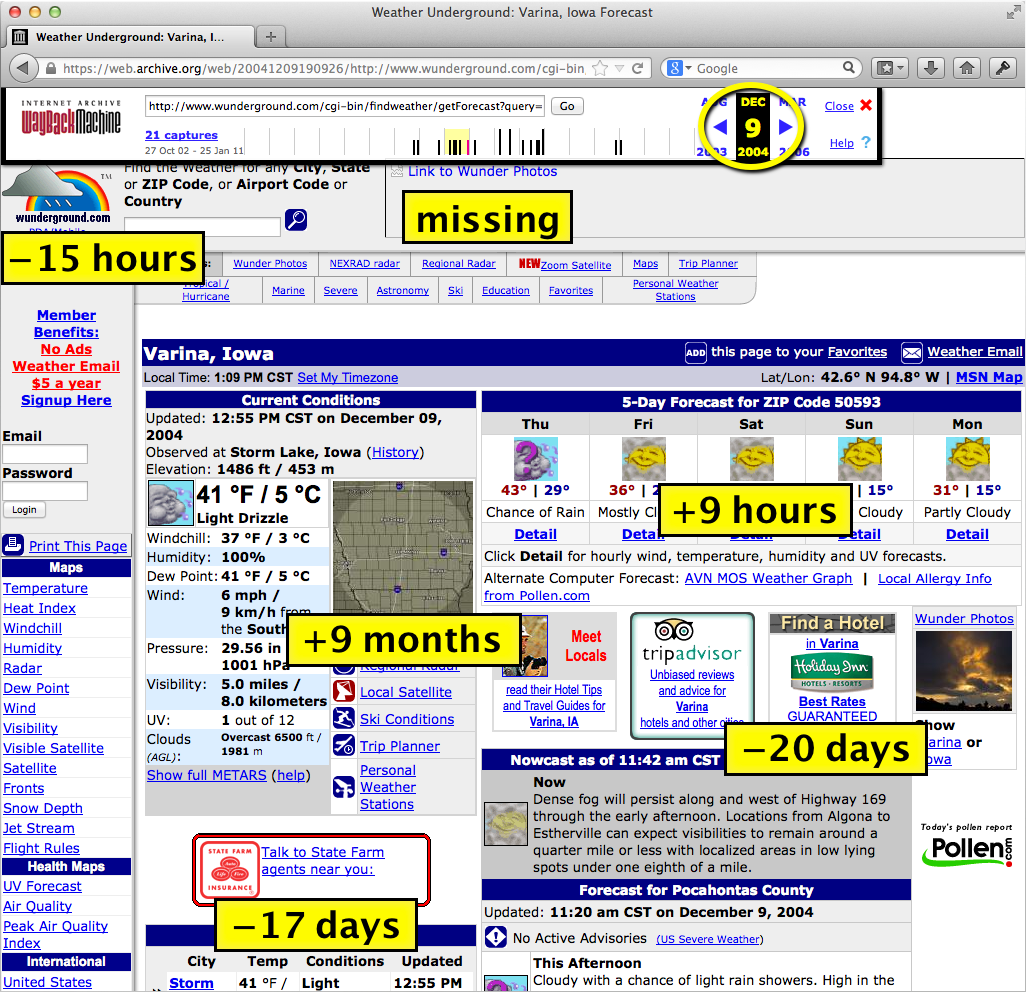}
  \caption{wunderground.org forecast page, Dec. 9, 2004}
  \label{f:webpage:plain}
\end{figure}

Temporal spread is the difference between the earliest and latest
  Memento-Datetimes in a composite memento.
Consider again the December 9, 2004 composite memento for the wunderground.org
  page, which is shown again in figure \ref{f:webpage:plain}.
It comprises a root memento\footnote{\url{http://web.archive.org/web/20041209190926/http://www.wunderground.org/cgi-bin/findWeather/getForecast?query=50593}}
  and 128 embedded resources, a sample of which are shown in table \ref{t:mementos}.
The earliest capture occurred 1.8 months before the root was captured.
The last capture occurred \textit{8.1 years} after the root was captured.
The temporal spread is 8.3 years, the mean delta is 1.8 years and the
  standard deviation is 2.9 years.
Thirty four of the embedded URI-Rs are not archived (e.g., the banner ad
  labeled missing in figure \ref{f:webpage:plain}) and one is archived
  (a timemap exists) but the desired embedded memento is not availble.

\begin{table*}
\centering
\caption{Embedded Memento Capture Datetimes (15 of 128)}
\label{t:mementos}
\renewcommand{\tabcolsep}{1mm}
\begin{tabularx}{\linewidth}{Xcr@{\ }l}
  URI & ~~~~~Memento-Datetime~~~~~ & \multicolumn{2}{c}{Delta} \\
  \hline
  http://ads.wunderground.com/ads/images/wu-9.jpg
    & 2004-10-16 03:40:53Z & -1.8 & months \\
  http://icons.wunderground.com/graphics/smalllogo2.gif
    & 2004-11-22 05:46:03Z &  -18~~~ & days \\
  http://ads.wunderground.com/ads/images/Statefarm-sfcom001\_120x30.gif
    & 2004-11-22 05:46:32Z &  -18~~~ & days \\
  http://icons.wunderground.com/ads/images/Davi-00009-vp2\_125x125.gif
    & 2004-12-08 07:01:08Z &   -2~~~ & days \\
  http://icons.wunderground.com/graphics/smash/wunderTransparent.gif
    & 2004-12-09 04:36:14Z &  -15~~~ & hours \\
  http://www.wunderground.com/cgi-bin/findweather/getForecast?query=50593
    & 2004-12-09 19:29:26Z &  --~~~~ & \textit{root} \\
  http://icons.wunderground.com/graphics/conds/cloudy.GIF
    & 2004-12-10 04:48:55Z &  +21~~~ & minutes \\
  http://icons.wunderground.com/graphics/conds/mostlycloudy.GIF
    & 2004-12-10 04:48:55Z &  +10~~~ & hours \\
  http://icons.wunderground.com/graphics/conds/rait.GIF
    & 2004-12-12 14:54:01Z &   +3~~~ & days \\
  http://icons.wunderground.com/ads/images/TripAdvisor-Blinky.gif
    & 2005-01-27 02:58:30Z &   +1.6 & months \\
  http://banners.wunderground.com/cgi-bin/statefarmbanner?zip=50593\&width=150
    & 2006-03-26 03:29:00Z &   +1.3 & years \\
  http://www.valueclick.com/system/files/coupon-mountain-slider.png?1310511429
    & 2011-07-13 09:08:00Z &   +6.6 & years \\
  http://z1.adserver.com/system/files/logo.gif
    & 2013-01-13 06:09:14Z &   +8.1 & years \\
  http://icons.wunderground.com/data/wximagenew/d/d70dave/0-thumb.jpg
    & Not Archived         &        &       \\
  http://pagead2.googlesyndication.com/pagead/show\_ads.js
    & Missing Memento      &        &       \\
  \hline
\end{tabularx}
\end{table*}

\subsection{Temporal Coherence}
\label{ss:TemporalCoherence}

We define an embedded memento to be temporally coherent with respect to
  a root memento when it can be shown that the embedded memento's representation
  existed at the time the root memento was captured.
So, which of the embedded mementos listed in table \ref{t:mementos}
  are temporally coherent?
The answer to this question requires evaluating the relationship between the
  root memento and the embedded memento.
This relationship can be one of many patterns, described in
  section \ref{s:TemporalCoherencePatterns}.

\section{Temporal Coherence Patterns}
\label{s:TemporalCoherencePatterns}

\newcommand{\isdef}[0]{\!\downarrow\!}
\newcommand{\isndef}[0]{\!\uparrow\!}
\newcommand{\body}[1]{\ensuremath{\mathcal{B}_{#1}}}
\newcommand{\lm}[1]{\ensuremath{\mathcal{L}_{#1}}}
\newcommand{\mdt}[1]{\ensuremath{\mathcal{T}_{#1}}}
\newcommand{\impC}{\implies \textrm{C}}
\newcommand{\impV}{\implies \textrm{V}}
\newcommand{\impPC}{\implies \textrm{PC}}
\newcommand{\impPV}{\implies \textrm{PV}}
\newcommand{\AND}{\,\wedge\,}

\newcommand{\oneRB}{\lm{i,1}\isdef \AND (\lm{i,1} \le \mdt{0} < \mdt{i,1})}
\newcommand{\oneRN}{\lm{i,1}\isdef \AND (\mdt{0} < \lm{i,1} \le \mdt{i,1})}
\newcommand{\oneRU}{\lm{i,1}\isndef \AND (\mdt{0} < \mdt{i,1})}
\newcommand{\oneLL}{\lm{i,n}\isdef \AND (\mdt{i,n} < \mdt{0})}
\newcommand{\oneLU}{\lm{i,n}\isndef \AND (\mdt{i,n} < \mdt{0})}
\newcommand{\oneEQ}{\mdt{0} = \mdt{i,j}}
\newcommand{\twoB}{\lm{i,j}\isdef \AND (\mdt{i,j-1} < \lm{i,j} < \mdt{0} < \mdt{i,1})}
\newcommand{\twoN}{\lm{i,1}\isdef \AND (\mdt{i,j-1} < \mdt{0} < \lm{i,1} < \mdt{i,1})}
\newcommand{\twoU}{\lm{i,1}\isndef \AND (\mdt{i,j-1} < \mdt{0} < \mdt{i,1})}
\newcommand{\twoBf}[1]{\twoB \AND (\emem{j-1} #1 \emem{j})}
\newcommand{\twoNf}[1]{\twoN \AND (\emem{j-1} #1 \emem{j})}
\newcommand{\twoUf}[1]{\twoU \AND (\emem{j-1} #1 \emem{j})}

\begin{figure}[b]
  \centering
  \includegraphics[width=\columnwidth]{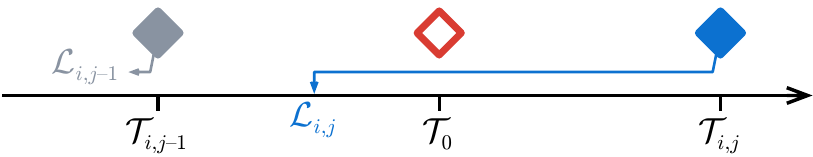}
  \caption{Sample Pattern Chart}
  \label{f:SamplePatternChart}
\end{figure}

The temporal coherence patterns defined below occurs are based on the
  relationship between \textrm{Memento-Datetime},
  \textrm{Last-Modified} datetime, and entity body content of a root and
  one or two embedded mementos.
Symbols for these attributes are defined in table \ref{t:attributes}.
The relationships are illustrated in charts like the one in figure
  \ref{f:SamplePatternChart}.
Each diamond represents a memento.
The red hollow diamond is the best memento for the root;
  solid diamonds are mementos for an embedded resource.
Blue objects affect the coherence state; gray objects do not.
The long horizontal black line is a time line, with earlier datetimes
  to the left as usual.
Mementos and \textrm{Last-Modified} datetimes are plotted relative to each other.
The time line is not scaled; the distance between datetimes is not meaningful.
\textrm{Last-Modified} datetimes are associated with mementos
  by a line pointing from the memento diamond to $\lm{}$.
Undefined data is denoted with an up arrow ($\uparrow$).
Likewise, a down arrow ($\downarrow$) asserts that an attribute is or must
  be defined.
(Arrow notation adopted from Kieffer et al.\ \cite{kieffer-arXiv:0805.1386}.)

\begin{table}[t]
\centering
\caption{Memento Attributes}
\label{t:attributes}
\renewcommand{\tabcolsep}{1mm}
\begin{tabularx}{\linewidth}{lX}
  Term & Definition \\
  \hline
  \mdt{0} & \textit{Memento-Datetime} of root memento \rmem. \\
  \lm{0}  & \textit{Last-Modified}\textsuperscript{\textdagger} datetime of
            root memento \rmem. \\
  \hline
  \mdt{i,j}  & \textit{Memento-Datetime} of embedded memento \emem{j}. \\
  \lm{i,j}   & \textit{Last-Modified}\textsuperscript{\textdagger} datetime of
               memento embedded \emem{j} \\
  \body{i,j} & Entity body\textsuperscript{\textdagger} of memento embedded \emem{j}. \\
  \hline
  \multicolumn{2}{l}{\footnotesize \textsuperscript{\textdagger}See
                     RFC 2616 \cite{rfc2616}.}
\end{tabularx}
\end{table}

\subsection{Temporal Coherence States}

Each pattern is categorized into one of five temporal coherence states:
\begin{description}
  \item[Prima Facie Coherent (C)]
    The embedded memento existed in its archived state at the time the root
    memento was captured.
  \item[Prima Facie Violative (V)]
    The embedded memento did not exist in its archived state at the time the
    root memento was captured.
  \item[Possibly Coherent (PC)]
    The embedded memento could have existed in its archive state at the time
    the root memento was captured.
  \item[Probably Violative (PV)]
    The embedded memento probably did not exist in its archived state at the
    time the root memento was captured.
  \item[Coherence Undefined (CU)]
    There is not enough information to determine coherence state.
\end{description}

\subsection{Pattern Groups}

The patterns are presented in four groups.
One-memento patterns consider a single memento for an embedded resource.
Two-memento patterns consider two mementos,
  one captured on or before the root and one captured after.
Content patterns are two-memento patterns that also consider memento
  content, not just capture and modification datetimes.
The other patterns group includes special cases such as a root
  memento that has no embedded resources.

\subsubsection{One-Memento Patterns}
\label{ss:OneMementoPatterns}

One-memento patterns consider one root URI-R, \rurir, and one embedded URI-R,
  \eurir{i}, and their corresponding mementos, \rmem and \emem{j} respectively.
If $\mdt{i,j} < \mdt{0}$ for all \emem{j}, then a
  \textit{left}\footnote{\label{fn:leftright}\textit{Left}
  and \textit{right} come from the appearance of the charts.}
  pattern exists and only the newest memento, \emem{n}, must be considered.
If $\mdt{i,j} \ge \mdt{0}$ for all \emem{j}, then a \textit{right}
  pattern exists and only the oldest memento, \emem{1}, must be considered.


\begin{figure}
  \centering
  \subfigure[Right Bracket (1RB) ~$\Rightarrow$~ C] {
	\includegraphics[width=0.98\columnwidth]{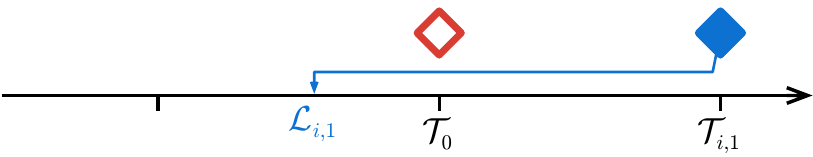}
    \label{f:oneRB}
  }
  \subfigure[Right Newer Last-Modified (1RN) ~$\Rightarrow$~ V] {
	\includegraphics[width=0.98\columnwidth]{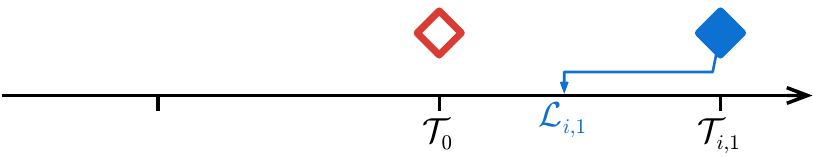}
    \label{f:oneRN}
  }
  \subfigure[Right Undefined Last-Modified (1RU) ~$\Rightarrow$~ PV] {
	\includegraphics[width=0.98\columnwidth]{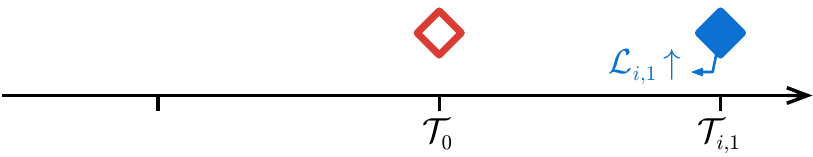}
    \label{f:oneRU}
  }
  \caption{Right-Sided Patterns}
  \label{f:RightSidedPatterns}
\end{figure}

\newcommand{\pattern}[1]{\textbf{\textit{#1}} \\}

\pattern{1RB: Right Bracket}
The right pattern depicted in figure \ref{f:oneRB}, and specified by
  predicate \eqref{e:oneRB}, represents an embedded memento, \emem{1}, which was
  captured after the root memento, \rmem, was captured, but was
  modified before \rmem was captured
The embedded memento's \textrm{Memento-Datetime} is after the root's
  and its \textrm{Last-Modified} datetime is on or before the root's
  \textrm{Memento-Datetime}.
The embedded memento's \textrm{Last-Modified} datetime and its
  \textrm{Memento-Datetime} bracket the root's \textrm{Memento-Datetime}.
Therefore, the embedded memento existed in its archived state at the time the
  root memento was captured.
Pattern 1RB is prima facie coherent.
\begin{equation}
  \oneRB \impC
  \label{e:oneRB}
\end{equation}

\pattern{1RN: Right Newer Last-Modified}
The right pattern depicted in figure \ref{f:oneRN}, and specified by
  predicate \eqref{e:oneRN}, represents an embedded memento, \emem{1}, which
  was both modified and captured after the root memento, \rmem, was captured.
The embedded memento's \textrm{Memento-Datetime} and \textrm{Last-Modified}
  datetime are both later than the root's \textrm{Memento-Datetime}.
This evidence indicates that the embedded memento was modified after the root
  memento; therefore, the embedded memento did not exist in its archived state
  at the time the root memento was captured.
Pattern 1RN is prima facie violative.
\begin{equation}
  \oneRN \impV
  \label{e:oneRN}
\end{equation}

\pattern{1RU: Right Undefined Last-Modified}
The right pattern depicted in figure \ref{f:oneRU}, and specified by
  predicate \eqref{e:oneRU}, represents an embedded memento, \emem{1}, which
  was captured after the root memento, \rmem, and does not have a
  \textrm{Last-Modified} datetime.
The embedded memento's \textrm{Memento-Datetime} is after the root's, like the
  1RN pattern, but in this case the \textrm{Last-Modified} datetime is undefined.
This evidence alone does not allow determination of the embedded memento's
  state at the time the root memento was captured.
However, since the embedded URI-R is referenced by the root memento, it is
  likely that a representation existed at the time the root was captured.
Our experience is that a missing \textrm{Last-Modified} datetime header
  normally indicates a dynamically-generated resource generated
  on demand (explanation in Appendix B).
Thus, pattern 1RU is probably violative.
\begin{equation}
  \oneRU \impPV
  \label{e:oneRU}
\end{equation}


\begin{figure}
  \centering
  \subfigure[Left Last-Modified (1LL) ~$\Rightarrow$~ PC] {
	\includegraphics[width=0.98\columnwidth]{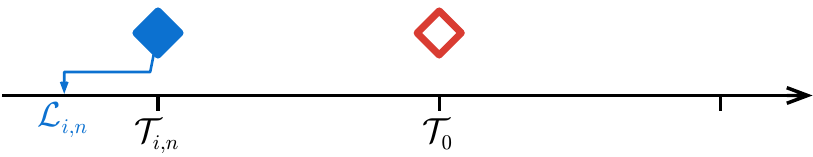}
    \label{f:oneLL}
  }
  \subfigure[Left Undefined Last-Modified (1LU) ~$\Rightarrow$~ PV] {
	\includegraphics[width=0.98\columnwidth]{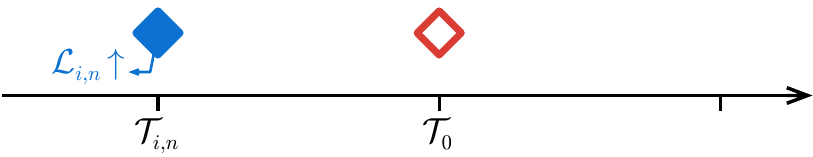}
    \label{f:oneLU}
  }
  \caption{Left-Sided Patterns}
  \label{f:LeftSidedPatterns}
\end{figure}

\pattern{1LL: Left Last-Modified}
The left pattern depicted in figure \ref{f:oneLL}, and specified by
  predicate \eqref{e:oneLL}, represents an embedded memento, \emem{n}, which
  was captured before the root memento, \rmem, and has a \textrm{Last-Modified}
  datetime.
Because the embedded memento's \textrm{Memento-Datetime} is before the root's,
  the \textrm{Last-Modified} datetime does not directly
  affect coherence state.
However, the existence of \textrm{Last-Modified} indicates that the embedded
  memento was probably not dynamically generated.
Therefore, pattern 1LL is probably coherent.
\begin{equation}
  \oneLL \impPC
  \label{e:oneLL}
\end{equation}

\pattern{1LU: Left Undefined Last-Modified}
The left pattern depicted in figure \ref{f:oneLU}, and specified by
  predicate \eqref{e:oneLU}, represents an embedded memento, \emem{n}, which
  was captured before the root memento, \rmem, and does not have a
  \textrm{Last-Modified} datetime.
The embedded memento's \textrm{Memento-Datetime} is before the root's,
  which means the lack of a \textrm{Last-Modified} datetime does
  not directly affect coherence state.
However, the lack of \textrm{Last-Modified} implies that
  the embedded resource was probably dynamically generated.
Therefore, pattern 1LU is probably violative.
\begin{equation}
  \oneLU \impPV
  \label{e:oneLU}
\end{equation}

\begin{figure}
  \centering
  \includegraphics[width=0.98\columnwidth]{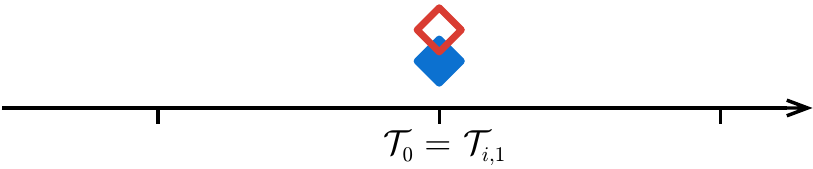}
  \caption{Simultaneous Capture ~$\Rightarrow$~ C}
  \label{f:oneEQ}
\end{figure}

\pattern{1EQ: Simultaneous Capture}
The pattern depicted in figure \ref{f:oneEQ}, and specified by
  predicate \eqref{e:oneEQ}, represents an embedded memento captured
  simultaneously with the root\footnote{Exact simultaneity is improbable.
  However, \textrm{Memento-Datetime} precision is one second and a web browser is
  unlikely to download multiple copies within one second. Therefore, the captures
  are effectively simultaneous.}.
Because the embedded memento's \textrm{Memento-Datetime} equals the root's,
  pattern 1EQ is prima facie coherent.
\begin{equation}
  \oneEQ \impC
  \label{e:oneEQ}
\end{equation}

\subsubsection{Two-Memento Patterns}
\label{ss:TwoMementoPatterns}

Two-memento patterns consider one root URI-R, \rurir, with a single memento,
  \rmem, and one embedded URI-R, \eurir{i}, with two consecutive mementos,
  \emem{j-1} and \emem{j}.
The consecutive mementos are selected such that
  $\mdt{i,j-1} < \mdt{0} < \mdt{i,j}$.
All three two-memento patterns share this characteristic.

\begin{figure}
  \centering
  \subfigure[Bracket (2B) ~$\Rightarrow$~ C] {
	\includegraphics[width=0.98\columnwidth]{figures/2B.pdf}
    \label{f:twoB}
  }
  \subfigure[Newer Last-Modified (2N) ~$\Rightarrow$~ V] {
	\includegraphics[width=0.98\columnwidth]{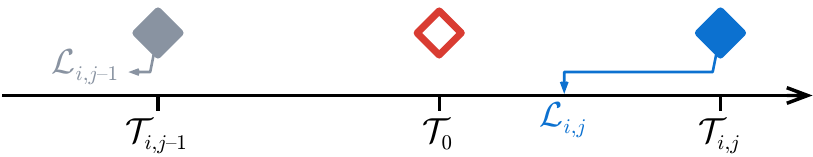}
    \label{f:twoN}
  }
  \subfigure[Undefined Last-Modified (2U) ~$\Rightarrow$~ PV] {
	\includegraphics[width=0.98\columnwidth]{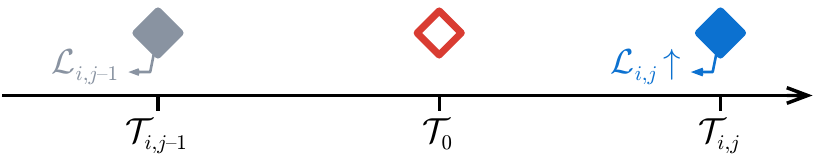}
    \label{f:twoU}
  }
  \caption{Two-Memento Patterns}
  \label{f:TwoMementoPatterns}
\end{figure}

The two-memento patterns are closely related to similar right 
  (1RB, 1RN, and 1RU) patterns.
The primary difference is the addition of a left memento to the
  two-memento patterns.
The two-memento patterns are named 2B, 2N, and 2U after their one-memento
  counterparts and are depicted in figure \ref{f:TwoMementoPatterns}.
Note that all the left mementos are shown in gray, which denotes
  that they do not affect coherence state.

\pattern{2B: Two-Memento Bracket}
The two-memento pattern depicted in figure \ref{f:twoB}, and specified by
  predicate \eqref{e:twoB}, represents a pair of mementos for an
  embedded resource,
  \emem{j-1} and \emem{j}, with \emem{j-1} captured before the root memento and
  \emem{j} captured after the root memento.
\emem{j} has a \textrm{Last-Modified} datetime that is on or before the
  root's capture time.
Thus, \emem{j} is under the same conditions in this pattern as it is in 1RB.
This can be seen by comparing \eqref{e:twoB} with \eqref{e:oneRB}.
Therefore, the embedded memento existed in its archived state at the time the
  root memento was captured and Pattern 2B is prima facie coherent.
(Note: Because \emem{j} brackets \rmem, \emem{j-1} does not affect temporal
  coherence state.)
\begin{equation}
  \twoB \impC
  \label{e:twoB}
\end{equation}

\pattern{2N: Two-Memento Newer Last-Modified}
The two-memento pattern depicted in figure \ref{f:twoN}, and specified by
  predicate \eqref{e:twoN}, represents a pair of mementos for an
  embedded resource,
  \emem{j-1} and \emem{j}, the same capture timing as in pattern 2B.
This pattern differs from 2B in that \emem{j} has a \textrm{Last-Modified}
  datetime that after root's capture time.
Thus, \emem{j} is under the same conditions in this pattern as it is in 1RN.
This can be seen by comparing \eqref{e:twoN} with \eqref{e:oneRN}.
Therefore, like pattern 2RN, pattern 2N is prima facie violative.
(Note: Because the state of \emem{j-1} is unknown from $\mdt{i,j-1}$ to
  $\lm{i,j}$, it does not affect temporal coherence state.)
\begin{equation}
  \twoN \impV
  \label{e:twoN}
\end{equation}

\pattern{2U: Two-Memento Undefined Last-Modified}

The two-memento pattern depicted in figure \ref{f:twoU}, and specified by
  predicate \eqref{e:twoU}, represents a pair of mementos for an
  embedded resource,
  \emem{j-1} and \emem{j}, the same capture timing as in patterns 2B and 2N.
This pattern differs from 2B and 2N in that \emem{j} has an undefined
  \textrm{Last-Modified} datetime.
Thus, \emem{j} is under the same conditions in this pattern as it is in 1RU.
This can be seen by comparing \eqref{e:twoU} with \eqref{e:oneRU}.
Therefore, like pattern 1RU, pattern 2U is probably violative.
\begin{equation}
  \twoU \impPV
  \label{e:twoU}
\end{equation}

\subsubsection{Content Patterns}
\label{ss:ContentPatterns}

Even when \textrm{Last-Modified} datetime is invalid or unavailable, when
  two mementos are available, additional evidence is available: the mementos'
  content, and more importantly, their similarity or lack thereof.

Content patterns are extensions to the two-memento patterns that add
  memento content to the determination of coherence state.
Content patterns require more computing resources and time than the
  one- and two-memento patterns because two memento entity
  bodies must be retrieved for each embedded URI-R.
This additional cost may render content patterns unsuitable for casual
  archive use or in restricted bandwidth conditions.

\begin{figure}[b!]
  \centering
  \subfigure[Bracket w/Evaluation (2EB, 2SB, 2NB) $\Rightarrow$ (C,C,C)] {
	\includegraphics[width=0.98\columnwidth]{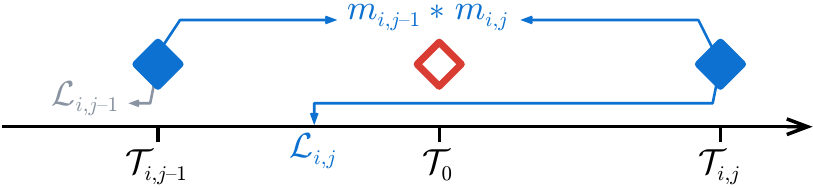}
    \label{f:twoBf}
  }
  \subfigure[Newer Last-Modified w/Content (2EN, 2SN, 2NN) $\Rightarrow$ (C,C,C)] {
	\includegraphics[width=0.98\columnwidth]{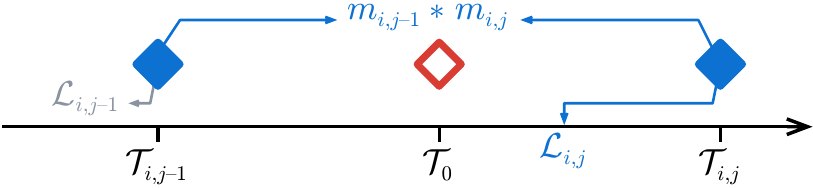}
    \label{f:twoNf}
  }
  \subfigure[Undef. Last-Modified w/Content (2EU, 2SU, 2NU) $\Rightarrow$ (C,C,V)] {
	\includegraphics[width=0.98\columnwidth]{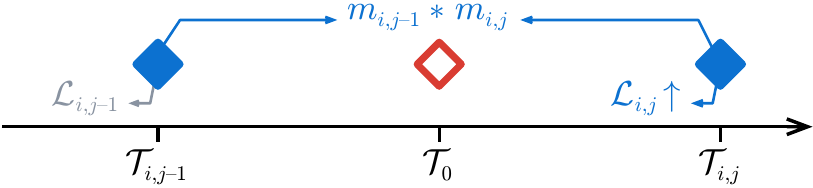}
    \label{f:twoUf}
  }
  \caption{Content Patterns}
  \label{f:ContentPatterns}
\end{figure}

The Content patterns are depicted in figure \ref{f:ContentPatterns}.
Each of the three charts represents a class of patterns.
These pattern classes are extensions of the two-memento patterns defined in
  \ref{ss:TwoMementoPatterns}, differing in how memento content is evaluated
  and how it affects coherence state.
These differences are represented by the asterisk operator ($\ast$), which
  in turn represents an evaluation function.
Evaluation functions provide evidence about the archival state of the embedded
  memento, \emem{j-1}, at the time the root memento, \rmem, was captured.
There are three evaluation functions:
\begin{description}
  \item[Equals ($=$)]
    $\body{\emem{j-1}}\!= \body{\emem{j}}$.
    This function returns true if the bodies of \emem{j-1} and \emem{j} are
    bit-for-bit equal.  A true value indicates that \emem{j-1} probably
    existed in its archived state at the time \rmem was captured.
  \item[Similar ($\sim$)]
    $\body{\emem{j-1}}\!\sim \body{\emem{j}}$.
    This function returns true if the bodies of \emem{j-1} and \emem{j} are
    substantially similar.
    A true value asserts with high confidence that \emem{j-1} existed in its
    archived state at the time the \rmem was captured.
  \item[Not Similar ($\nsim$)]
    $\body{\emem{j-1}}\!\nsim \body{\emem{j}}$.
    This function returns true if the bodies of \emem{j-1} are not
    substantially similar.
    Note, for mementos, $m_i \nsim m_j \implies m_i \ne m_j$.
\end{description}

The reason for both equality and similarity is that many web archives treat
  text resources (e.g., HTML) and binary resources (e.g., images) differently.
Archive metadata is frequently added to text resources, while binary resources
  are not changed.
It makes sense to first check equality, and fall back to similarity
  if the equality check fails.
It should be noted that while the definition equality is universal,
  the definition of similar will vary by application and user need.

All of the equality and similarity patterns result
  in the Prima Facie Coherent temporal coherence state.
This raises the question of whether or not all six patterns are required.
Although a single equality pattern and single similarity pattern are
  sufficient, we have chosen to retain all six patterns for the present.

\pattern{2EB: Content Equal Bracket}
The content pattern depicted in figure \ref{f:twoBf}, and specified by
  predicate \eqref{e:twoBE}, is the same as pattern 2B plus the determination
  that the two embedded mementos have bit-for-bit equality.
Bit-for-bit equality is an overarching condition, which when combined with the
  \textrm{Last-Modified} datetime evidence described in patterns 1RB and 2B,
  provides strong evidence the embedded memento existed in its archived state
  at the time the root memento was captured.
Thus, pattern 2EB is prima facie coherent.
\begin{equation}
  \twoBf{=} \impC
  \label{e:twoBE}
\end{equation}

\pattern{2EN: Content Equal Newer Last-Modified}
The content pattern depicted in figure \ref{f:twoNf}, and specified by
  predicate \eqref{e:twoNE}, is the same as pattern 2N plus the determination
  that the two embedded mementos have bit-for-bit equality.
Bit-for-bit equality is an overarching condition which provides evidence that
  the embedded memento existed in its archived state at the time the root
  memento was captured; it overrides the \textrm{Last-Modified} datetime
  evidence described in related patterns 1RN and 2N.
Thus, pattern 2EN is prima facie coherent.
\begin{equation}
  \twoNf{=} \impC
  \label{e:twoNE}
\end{equation}

\pattern{2EU: Content Equal Undefined Last-Modified}
The content pattern depicted in figure \ref{f:twoUf}, and specified by
  predicate \eqref{e:twoUE}, is the same as pattern 2U plus the determination
  that the two embedded mementos have bit-for-bit equality.
Bit-for-bit equality is an overarching condition which provides evidence that
  the embedded memento existed in its archived state at the time the root
  memento was captured; it overrides the undefined \textrm{Last-Modified}
  datetime evidence described in related patterns 1RU and 2U.
Thus, pattern 2EU is prima facie coherent.
\begin{equation}
  \twoUf{=} \impC
  \label{e:twoUE}
\end{equation}

\pattern{2SB: Content Similar Bracket}
The content pattern depicted in figure \ref{f:twoBf}, and specified by
  predicate \eqref{e:twoBS}, is a weaker form of pattern 2EB.
In place of bit-for-bit equality, a similarity measure it used.
Like equality, sufficient similarity is an overarching condition and provides
  evidence that the embedded memento existed in its archived state at the time
  the root memento was captured; it overrides the \textrm{Last-Modified}
  datetime evidence described in related patterns 1RB and 2B.
Thus, pattern 2SB is prima facie coherent.
\begin{equation}
  \twoBf{\sim} \impC
  \label{e:twoBS}
\end{equation}

\pattern{2SN: Content Similar Newer Last-Modified}
The content pattern depicted in figure \ref{f:twoNf}, and specified by
  predicate \eqref{e:twoNS}, a weaker form of pattern 2EN.
Like pattern 2SB, a similarity measure is used in place of bit-for-bit equality.
Like equality, sufficient similarity is an overarching condition and provides
  evidence that the embedded memento existed in its archived state at the time
  the root memento was captured; it overrides the \textrm{Last-Modified}
  datetime evidence described in related patterns 1RN and 2N.
Thus, pattern 2EN is prima facie coherent.
\begin{equation}
  \twoNf{\sim} \impC
  \label{e:twoNS}
\end{equation}

\pattern{2SU: Content Similar Undefined Last-Modified}
The content pattern depicted in figure \ref{f:twoUf}, and specified by
  predicate \eqref{e:twoUS}, a weaker form of pattern 2EU.
Like patterns 2SB and 2SN, a similarity measure is used in place of bit-for-bit
  equality.
Like equality, sufficient similarity is an overarching condition and provides
  evidence that the embedded memento existed in its archived state at the time
  the root memento was captured.
This similarity overrides the \textrm{Last-Modified} datetime
  evidence described in related patterns 1RU and 2U.
\begin{equation}
  \twoUf{\sim} \impC
  \label{e:twoUS}
\end{equation}

\pattern{2NB: Content Not Similar Bracket}
The content pattern depicted in figure \ref{f:twoBf}, and specified by
  predicate \eqref{e:twoBN}, is the same as pattern 2B plus the determination
  that the two embedded mementos have different content.
Unlike the substantial similarity of pattern 2EB, the lack of similarity is
  not an overarching condition and does not add significant evidence that
  the embedded memento existed in its archived state at the time the root
  memento was captured.
Pattern 2NB is therefore equivalent to pattern 2B, which is prima facie coherent.
\begin{equation}
  \twoBf{\nsim} \impC
  \label{e:twoBN}
\end{equation}

\pattern{2NN: Content Not Similar Newer Last-Modified}
The content pattern depicted in figure \ref{f:twoNf}, and specified by
  predicate \eqref{e:twoNN}, is the same as pattern 2N plus the determination
  that the two embedded mementos have different content.
Unlike the substantial similarity of pattern 2EN, the lack of similarity is
  not an overarching condition and does not add significant evidence that
  the embedded memento existed in its archived state at the time the root
  memento was captured.
Pattern 2NN is therefore equivalent to pattern 2N, which is prima facie violative.
\begin{equation}
  \twoNf{\nsim} \impV
  \label{e:twoNN}
\end{equation}

\pattern{2NU: Content Not Similar Undefined Last-Modified}
The content pattern depicted in figure \ref{f:twoUf}, and specified by
  predicate \eqref{e:twoUN}, is the same as pattern 2U plus the determination
  that the two embedded mementos have different content.
Unlike the substantial similarity of pattern 2EU, the lack of similarity is
  not an overarching condition and does not add significant evidence that
  the embedded memento existed in its archived state at the time the root
  memento was captured.
Pattern 2NU is therefore equivalent to pattern 2U, which is probably violative.
\begin{equation}
  \twoUf{\nsim} \impPV
  \label{e:twoUN}
\end{equation}

\subsubsection{Other Patterns}
\label{ss:OtherPatterns}

There are several patterns that consider one root URI-R, \rurir, as do the
  one- and two-memento patterns, but do not consider embedded mementos.
The lack of embedded mementos means these patterns are either prima facie
  coherent or have undefined coherence state.

\begin{figure}
	\centering
    \subfigure[No Embedded URIs (0NE) ~$\Rightarrow$~ C] {
	  \includegraphics[width=0.98\columnwidth]{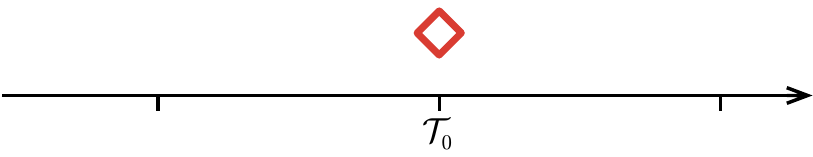}
	  \label{f:zeroNE}
    }
    \subfigure[Not Archived (0NA) ~$\Rightarrow$~ UC] {
	  \includegraphics[width=0.98\columnwidth]{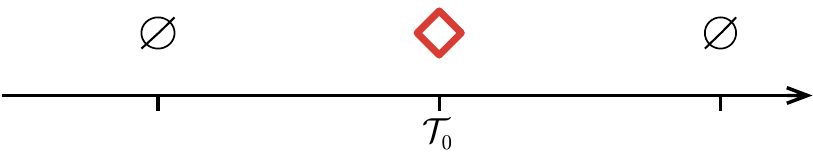}
	  \label{f:zeroNA}
    }
  \caption{Other Patterns}
  \label{f:OtherPatterns}
\end{figure}

\pattern{0NE: No Embedded URIs}
The pattern depicted in figure \ref{f:zeroNE} represents a root memento, \rmem,
  which has no embedded URIs (e.g., images, plain text).
These root mementos have inherent temporal coherence.

\pattern{0NA: Not Archived}
The pattern depicted in figure \ref{f:zeroNA} represents a root memento, \rmem,
  and an embedded URI-R, \eurir{i}, which is not archived (or the mementos are not
  available).
Temporal coherence state for this pattern is undefined.

\section{Web Data Is Imperfect}

The pattern descriptions in subsections
  \ref{ss:OneMementoPatterns}--\ref{ss:ContentPatterns} do not address
  the vast amounts of invalid and badly structured data
  present on the Web; and, consequently part of Web archive holdings.
This section describes a few of the imperfections we routinely encounter.

\subsection{Invalid Datetimes}
Datetimes in particular are critical to the identification of patterns and
  proper classification of coherence state.
In our work to date, every effort is made to \textit{decode} invalid datetimes.
Table \ref{t:DatetimeCorrections} lists the major corrective actions
  currently used.
If a datetime cannot be parsed, it is treated as undefined.

\begin{table}[t]
\centering
\caption{Datetime Corrections}
\label{t:DatetimeCorrections}
\renewcommand{\tabcolsep}{1mm}
\begin{tabularx}{\linewidth}{lX}
  Symptom & Description and Corrective Action \\
  \hline
  2-Digit year &
    This is symptomatic of correctly functioning pre-year 2000 software.
    Corrected by adding 1900 to the year. \\
  1-Digit year &
    This Y2K bug produced the correct 2-digit year prior to 2000, but fails
    from 2000 on. We have seen this error on datetimes as late as 2003.
    Corrected by adding 2000 to the year. \\
  3-Digit year &
    Another Y2K bug caused by subtracting 1900 to produce a 2-digit year.
    We have seen this error on datetimes as late as 2002.
    Corrected by adding 1900 to the year. \\
  Non-GMT &
    Local time zones are common in mementos captured before 2000.
    Corrected by converting to GMT. \\
  Non-English &
    Most datetime parsers only handle English month and weekday names, yet
    other languages crop up in datetime strings. The most common is French.
    Corrected using a lookup table to translate to English. \\
  Missing zeros &
    Missing leading zeros are common. Most common is day of month.
    Time fields are also affected. \\
  Extra zeros &
    Although not as common as missing leading zeros, extra leading zeros are
    found. Corrected by ignoring extra leading zeros. \\
  \hline
\end{tabularx}
\end{table}

\subsection{Last-Modified Datetimes}

Clausen \cite{clausen-iwaw04} demonstrated that Last-Modified datetimes
  are a reliable indicator of entity body change.
Therefore, we treat most Last-Modified datetimes as valid.
Clearly incorrect Last-Modified datetimes are treated as undefined
  (for example, when the original resource Last-Modified datetime is greater
  than the Memento-Datetime).
The two basic cases are (1) when the original resource server sets
  \textrm{Last-Modified} to the current datetime on every response, and
  (2) when the endity body is modified but original resource \textrm{Last-Modified}
  is not.
Both render attempts to determine coherence ineffective; fortunately, both
  are rare.
Much more common is the original resources server omitting
  \textrm{Last-Modified} altogether (patterns 1RU, 1LU, and 2U).
The content patterns defined in \ref{ss:ContentPatterns} (2EU, 2SU, and 2NU in
  particular) address the ommission using entity body equality and similarity.

It must also be noted that there are \textrm{pathological} cases caused by
  misconfigured servers and buggy software, where changes to the memento
  content and changes to \textrm{Last-Modified} datetime are unrelated.

\subsection{Missing Mementos}

Many of the patterns rely on retrieving memento HTTP headers and content.
However, this is not always possible.
The causes are not important, but the consequences must be considered.
For example, two consecutive mementos, \emem{j-1} and \emem{j}, would normally
  result in a two-memento pattern match.
But what if \emem{j-1} cannot be retrieved?
Should an \emem{j-2} be used as substitute?
Should a one-memento pattern be used instead?
These are open questions, the answers to which may depend on circumstances
  and the user priorities.

\subsection{Memento-Datetime Collisions}

Occasionally, an archive will have multiple mementos with the same
  \textrm{Memento-Datetime}.
There are many causes; three are briefly described here.
First, several archives can capture the same original
  resource at the same time.
If a client later uses multiple archives simultaneously, multiple
  mementos with the same \textrm{Memento-Datetime} will be found.
Second, a single capture is available from multiple archives
  because the archives share holdings
  (e.g., Archive-It! and the Internet Archive).
In both of these cases, the content is likely to be identical; however,
  metadata differences may exist (e.g., one archive captures
  \textrm{Last-Modifed} datetime and the other does not).
It is therefore possible for more than one pattern to be matched.
In this case, we suggest using the least favorable coherence state.
However, like the missing mementos issue, the best response to this condition
  may depend on user priorities.

A third scenario that leads to \textrm{Memento-Datetime} collisions is typified
  by Wikipedia edit wars \cite{wikipedia-editwar}.
In this case, changes occur less than one second apart.
Because \textrm{Memento-Datetime} inherits one-second resolution from HTTP,
  the resulting mementos can have the same \textrm{Memento-Datetime}.

\subsection{Redirection to the Live Web}

The Internet Archive Wayback Machine makes every effort to fulfill requests
  for embedded mementos.
In some cases this means redirecting to a Live Web resource.
In our research, we detect this redirection and treat the URI-R as if
 it has no mementos.

\section{Conclusion}

We have introduced temporal coherence patterns of embedded mementos and
four coherence states: prima facie coherent, possibly coherent,
  probably violative, and prima facie violative.
The patterns and resulting states, are summarized in table
  \ref{t:CoherencePatternsSummary},
Together, the patterns and states provide a framework in which to examine
  the temporal coherence of embedded mementos.

\begin{table*}
\centering
\caption{Coherence Patterns Summary}
\label{t:CoherencePatternsSummary}
\renewcommand{\tabcolsep}{1mm}
\begin{tabular*}{\textwidth}{
    @{\extracolsep{\fill}}
    @{\hspace{1mm}} llll @{\hspace{1mm}}}
  Abbr. & Description & Condition & Coherence State \\

  \hline\noalign{\smallskip}
  \multicolumn{4}{l}{\textbf{One-Memento Patterns}} \\
  \noalign{\smallskip}

  1RB & Right Bracket                 & $\oneRB$ & Prima Facie Coherent \\
  1RN & Right Newer Last-Modified                   & $\oneRN$ & Prima Facie Violative \\
  1RU & Right Undefined Last-Modified & $\oneRU$ & Probably Violative \\
  1LL & Left Last-Modified            & $\oneLL$ & Possibly Coherent \\
  1LU & Left Undefined Last-Modified  & $\oneLU$ & Probably Violative \\
  1EQ & Simultaneous Capture          & $\oneEQ$ & Prima Facie Coherent \\

  \hline\noalign{\smallskip}
  \multicolumn{4}{l}{\textbf{Two-Memento Patterns}} \\
  \noalign{\smallskip}

  2B & Bracket                 & $\twoB$ & Prima Facie Coherent \\
  2N & Newer Last-Modified                   & $\twoN$ & Prima Facie Violative \\
  2U & Undefined Last-Modified & $\twoU$ & Probably Violative \\

  \hline\noalign{\smallskip}
  \multicolumn{4}{l}{\textbf{Content Patterns}} \\
  \noalign{\smallskip}

  2EB & Equal Bracket                 & $\twoBf{=}$ & Prima Facie Coherent \\
  2EN & Equal Newer Last-Modified     & $\twoNf{=}$ & Prima Facie Coherent \\
  2EU & Equal Undefined Last-Modified & $\twoUf{=}$ & Prima Facie Coherent \\

  2SB & Similar Bracket                 & $\twoBf{\sim}$ & Prima Facie Coherent \\
  2SN & Similar Newer Last-Modified     & $\twoNf{\sim}$ & Prima Facie Coherent \\
  2SU & Similar Undefined Last-Modified & $\twoUf{\sim}$ & Prima Facie Coherent \\

  2NB & Not Similar Bracket                 & $\twoBf{\nsim}$ & Prima Facie Coherent \\
  2NN & Not Similar Newer Last-Modified     & $\twoNf{\nsim}$ & Prima Facie Violative \\
  2NU & Not Similar Undefined Last-Modified & $\twoUf{\nsim}$ & Probably Violative \\

  \hline\noalign{\smallskip}
  \multicolumn{4}{l}{\textbf{Other Patterns}} \\
  \noalign{\smallskip}

  0EM & No Embedded URI-Rs & N/A -- intrinsically coherent & Prima Facie Coherent \\
  0NA & No Mementos for URI-R & N/A -- resource not archived & Undefined Coherence \\

  \hline\noalign{\smallskip}
\end{tabular*}
\end{table*}


\bibliographystyle{plain}

\section*{Acknowledgments}

This work supported in part by the NSF (IIS 1009392) and the Library of
  Congress.
We are grateful to the Internet Archive for their continued support of
  Memento access to their archive.
Memento is a joint project between the Los Alamos National Laboratory
  Research Library and Old Dominion University.

I also wish to extend thanks to Yorick Chollet, an intern with Los Alamos
  National Laboratories, who found cut and paste induced errors in the
  second version.
Predicates 13, 14, 16, 17, 19, and 20 are correct as of version 3.

\section*{Appendix A: {Last-Modified} Is Not {Memento- \\
Datetime}}

\begin{figure*}[b]
  \centering
  \includegraphics[width=\textwidth]{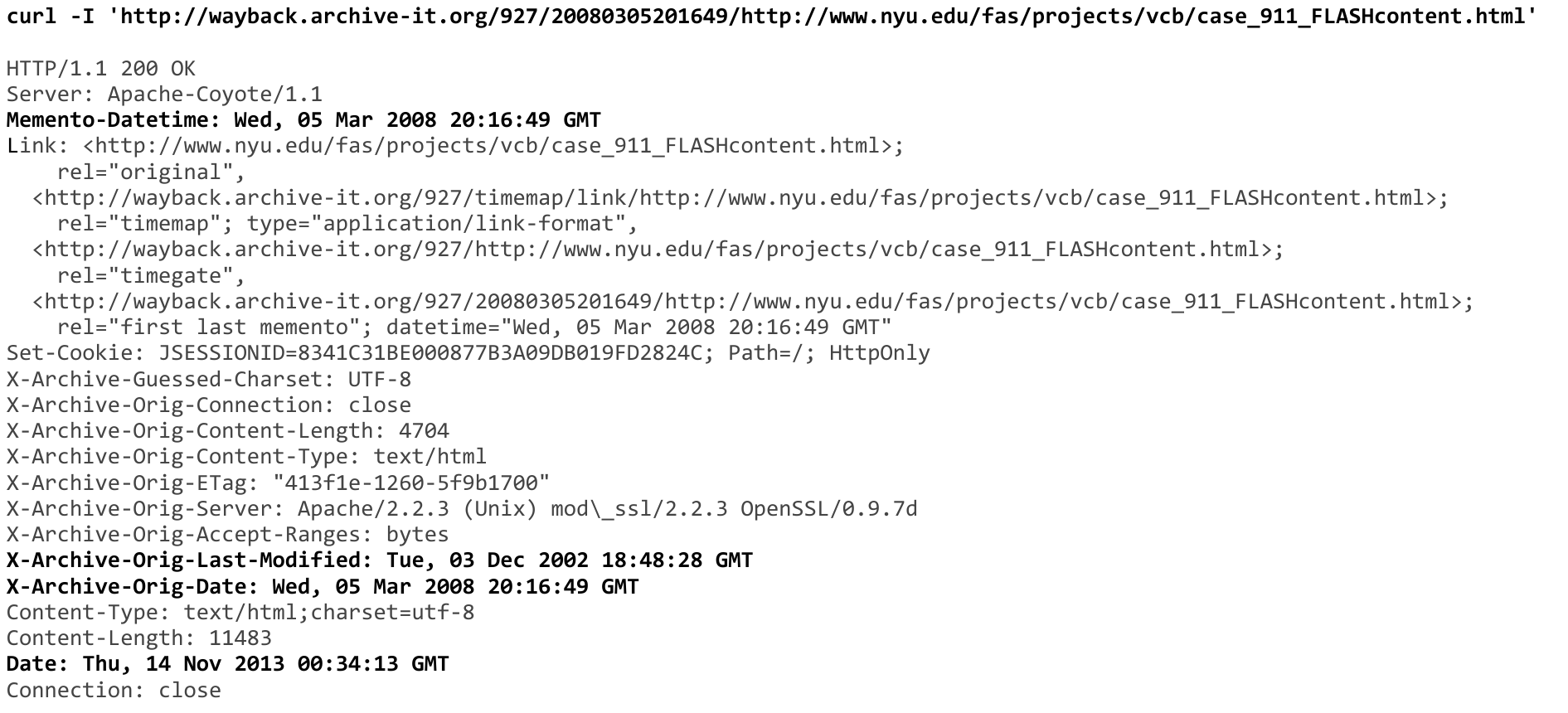}
  \caption{Memento Headers for a Presumably Statically-Generated Representation}
  \label{f:memento:headers:static}
\end{figure*}

Although similar at first look, the \textit{Memento-Datetime} and
  the \textit{Last-Modified} headers are not equivalent.
This appendix briefly describes the difference; we detail the difference
  in our blog post \cite{nelson-blog10.11.05},
Although, both are an indicator of when the state of a resource was known,
  Last-Modified is set by the original server and indicates when the
  resource was last modified, whereas Memento-Datetime is set by the
  archive server and indicates when the memento as captured.
Consider the URI in the blog post's second case (CD == MD < LM).
Deferencing it yields the headers shown in figure \ref{f:memento:headers:static}.
There are four datetime headers, which are indicated in bold.

Compare \textit{Date} to \textit{X-Archive-Original-Date}.
The Date header is the datetime the memento was dereferenced---the datetime
  the archive server responded to the \textit{curl} command.
X-Archive-Original-Date, on the other hand, is the datetime that the original
  resource was dereferenced---the datetime that the original server responded
  to the archive crawler's request.
Assuming that the original resource server and archive crawler have
  properly-synchronized clocks, X-Archive-Original-Date and
  Memento-Datetime will be identical or differ by only a second or two.
Both Date and Memento-Datetime indicate that the archive crawler
  captured the memento about 5.5 years ago.

Now look at \textit{X-Archive-Original-Last-Modified}, which is the
  Last-Modified header provided by the original resource server at the
  time the crawler captured the memento.
At the capture time in March 2008, the original resource server asserted that
  the page had not changed since December 2002.
When original resource servers are functioning properly and Last-Modified
  is provided, Last-Modified will precede Memento-Datetime, which will
  precede Date as shown in \eqref{e:lminmd:0}.
\begin{equation}
  \textrm{Last-Modified} < \textrm{Memento-Datetime} < \textrm{Date}
  \label{e:lminmd:0}
\end{equation}
For original resource servers that set Last-Modified to the current time,
  Last-Modified will equal Memento-Datetime differ by a second or two,
  as shown in \eqref{e:lminmd:1}.
\begin{equation}
  \textrm{Last-Modified} \approx \textrm{Memento-Datetime} < \textrm{Date}
  \label{e:lminmd:1}
\end{equation}

\section*{Appendix B: {Last-Modified} and Dynamic Representations}

\begin{figure*}[b]
  \centering
  \includegraphics[width=\textwidth]{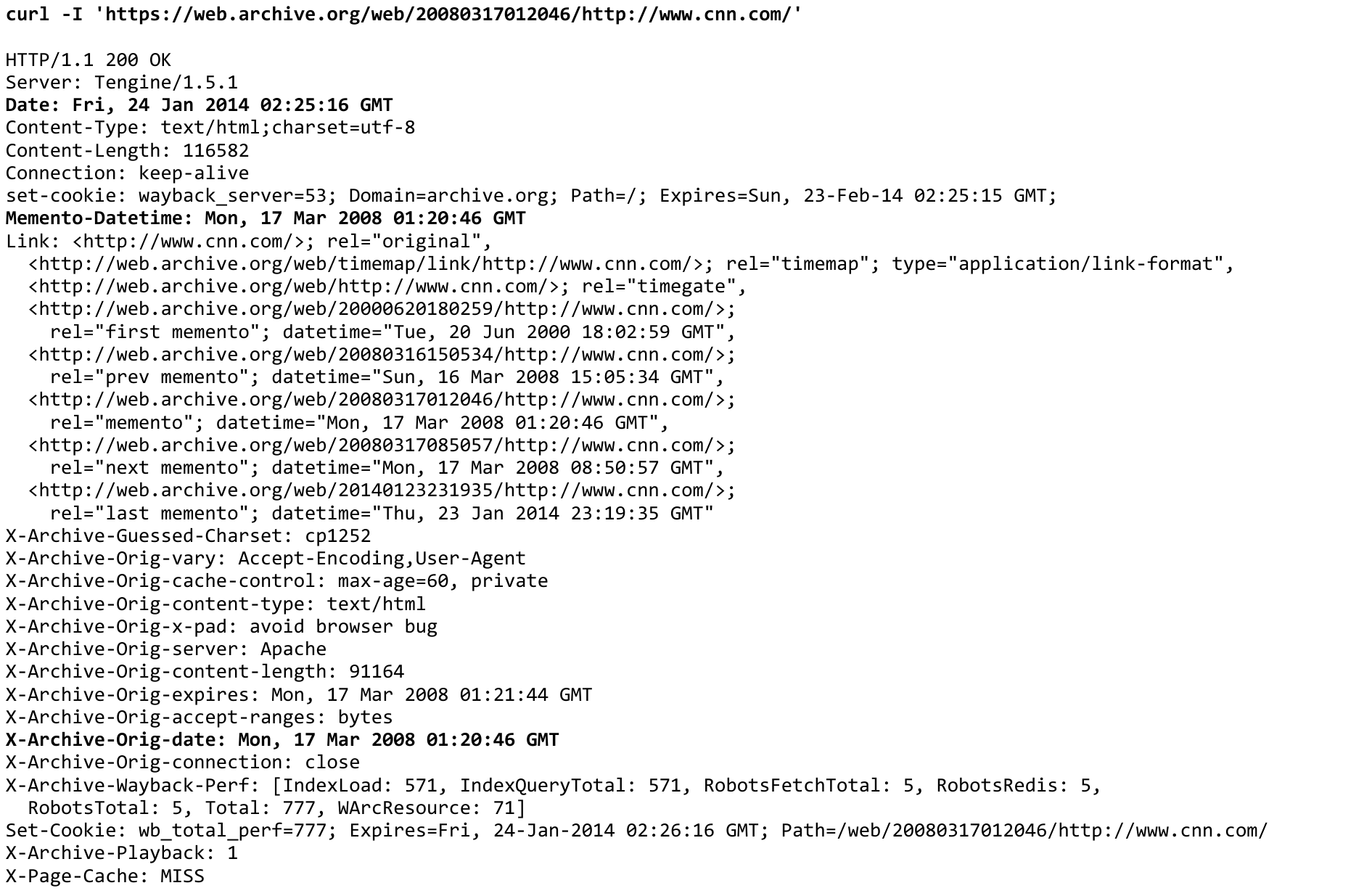}
  \caption{Memento Headers for a Presumably Dynamically-Generated Representation}
  \label{f:memento:headers:dynamic}
\end{figure*}

Although dynamically-generated representations can produce a
  \textrm{Last-Modified} header, in our experience they typically do not.
Also, when a \textrm{Last-Modified} header is produced for a dynamic
  representation, it is usually set to the current time.
Thus, we consider both the absence of \textrm{Last-Modified} or
  \textrm{Last-Modified} approximately equal to the \textrm{Date} header
  to be indicative of a dynamic representation.
Likewise, the presence of \textrm{Last-Modified} less than \textrm{Date}
  is indicative of a static representation.
Figure \ref{f:memento:headers:dynamic} shows the headers returned for the
  CNN home page, which is dynamically generated.
Unlike figure \ref{f:memento:headers:static} from Appendix A, the CNN home page
  lacks a \textit{X-Archive-Orig-Last-Modified} header.

We expect to empirically test the above assumptions in future work.
The end result is expected to be a model using \textrm{Last-Modified}, and
  possibily other headers, to determine the probability that an archived
  representation existed at the time the root memento was captured.

Note that the \textrm{ETag} header is defined to have similar, but more
  flexible, sematics than \textrm{Last-Modified} and can occur independently,
  they typically co-occur.
Thus, \textrm{ETag} generally adds little value in determining static or
  dynamic representation generation.

\end{document}